\documentclass[showpacs, secnumarabic, amssymb, nobibnotes, aip, jap, preprint]{revtex4-1}
\usepackage{graphicx}%
\usepackage{dcolumn}%
\usepackage{bm}%
\usepackage{lineno}

\begin{document}

\title{Magnetic Interactions in Ge$_{1\textrm{-}x}$Cr$_{x}$Te Semimagnetic Semiconductors}

\author{L.~Kilanski}
\email[Electronic mail: ]{kilan@ifpan.edu.pl}
\author{A.~Podg\'{o}rni}
\author{W.~Dobrowolski}
\author{M.~G\'{o}rska}
\author{B.~J.~Kowalski}
\author{A.~Reszka}
\author{V.~Domukhovski}
\author{A.~Szczerbakow}
\affiliation{Institute of Physics, Polish Academy of Sciences, al. Lotnikow 32/46, 02-668 Warsaw, Poland}

\author{K.~Sza{\l}owski}
\affiliation{Department of Solid State Physics, University of {\L}\'{o}d\'{z}, ul. Pomorska 149/153, 90-236 {\L}\'{o}d\'{z}, Poland}

\author{J.~R.~Anderson}
\author{N.~P.~Butch}
\altaffiliation[Present address: ]{Lawrence Livermore National Laboratory, 7000 East Avenue, Livermore, CA 94550, USA}
\affiliation{Department of Physics and Center for Nanophysics and Advanced Materials, University of Maryland, College Park, MD 20742, USA}

\author{V.~E.~Slynko}
\author{E.~I.~Slynko}
\affiliation{Institute of Materials Science Problems, Ukrainian Academy of Sciences, 5 Wilde Street, 274001 Chernovtsy, Ukraine}

\date{\today}

\begin{abstract}

We present the studies of magnetic properties of Ge$_{1\textrm{-}x}$Cr$_{x}$Te diluted magnetic semiconductor with changeable chemical composition 0.016$\,$$\leq$$\,$$x$$\,$$\leq$$\,$0.061. A spin-glass state (at $T$$\,$$\leq$$\,$35$\;$K) for $x$$\,$$=$$\,$0.016 and 0.025 and a ferromagnetic phase (at $T$$\,$$<$$\,$60$\;$K) for $x$$\,$$\geq$$\,$0.030 are observed. The long range carrier-mediated magnetic interactions are found to be responsible for the observed magnetic ordering for $x$$\,$$<$$\,$0.045, while for $x$$\,$$\geq$$\,$0.045 the spinodal decomposition of Cr ions leads to a maximum and decrease of the Curie temperature, $T_{C}$, with increasing $x$. The calculations based on spin waves model are able to reproduce the observed magnetic properties at a homogeneous limit of Cr alloying, e.g. $x$$\,$$<$$\,$0.04, and prove that carrier mediated Ruderman-Kittel-Kasuya-Yosida (RKKY) interaction is responsible for the observed magnetic states. The value of the Cr-hole exchange integral, $J_{pd}$, estimated via fitting of the experimental results with the theoretical model, is in the limits 0.77$...$0.88$\;$eV.

\end{abstract}





\maketitle

\linenumbers

\section{Introduction}

Diluted magnetic semiconductors (DMS) are a very important and intensively studied group of materials because of their possible application in spin electronics devices.\cite{Ohno10a, Dietl10a} The combination of magnetic ions and a semiconductor matrix allows an independent control of electrical and magnetic properties of DMS by many orders of magnitude via changes in the technological parameters of the growth or post growth treatment of the compound. DMS compounds are usually developed on the basis of a III-V or II-VI semiconductor matrix into which transition metal or rare earth ions are introduced on a level of several atomic percent.\cite{Dobrowolski03a, Matsukura02a, Kossut93a} \\ \indent IV-VI based DMS, in particular Ge$_{1\textrm{-}x}$TM$_{x}$Te alloys (TM - transition metal), possess many advantages over widely studied Ga$_{1\textrm{-}x}$Mn$_{x}$As. The carrier concentration and the amount of TM ions can be controlled independently. Moreover, the solubility of TM ions in GeTe is very high allowing growth of homogeneous Ge$_{1\textrm{-}x}$TM$_{x}$Te solid solutions over a wide range of chemical composition. Itinerant ferromagnetism with the Curie temperatures as high as 167$\;$K for $x$$\,$$=$$\,$0.5 was observed in bulk Ge$_{1\textrm{-}x}$Mn$_{x}$Te crystals.\cite{Cochrane74a} Recently, numerous papers were devoted to the studies of thin epitaxial Ge$_{1\textrm{-}x}$Mn$_{x}$Te layers.\cite{Hassan11a, Knoff09a, Fukuma08a} Recent results show that the Curie temperature in Ge$_{1\textrm{-}x}$Mn$_{x}$Te layers can be controlled in a wide range of values reaching a maximum of about 200$\;$K.\cite{Lechner10a} Apart from Mn-alloyed GeTe, a significant attention has been recently turned onto other transition-metal-alloyed GeTe based DMS,\cite{Fukuma03a} since they also show itinerant ferromagnetism with relatively high Curie temperatures $T_{C}$$\,$$\leq$$\,$150$\;$K.\cite{Fukuma03b, Fukuma06a, Fukuma06b} The progress in the technology of growth of a Ge$_{1\textrm{-}x}$Cr$_{x}$Te alloy resulted in obtaining a high Curie temperature reaching maximum of 180$\;$K for $x$$\,$$\approx$$\,$0.06.\cite{Fukuma07a} In contrast to the epitaxial layers, there seems to be no detailed study of magnetic behavior of bulk Ge$_{1\textrm{-}x}$Cr$_{x}$Te crystals. Therefore, detailed studies of Cr-alloyed GeTe crystals are very important in making further development in this class of compounds. \\ \indent In this paper we present studies of magnetic properties of Ge$_{1\textrm{-}x}$Cr$_{x}$Te bulk crystals with chemical composition $x$ varying between 0.016$\,$$\leq$$\,$$x$$\,$$\leq$$\,$0.061. Our main goal was to show that the magnetic properties of the alloy can be tuned by means of changes in its chemical composition. The present work extends our preliminary studies (see Ref.$\;$\onlinecite{Kilanski11a}) of low Cr-content Ge$_{1\textrm{-}x}$Cr$_{x}$Te samples with $x$$\,$$\leq$$\,$0.025 into higher Cr composition 0.035$\,$$\leq$$\,$$x$$\,$$\leq$$\,$0.061, in which the magnetic order changed dramatically from a spin-glass freezing into ferromagnetic alignment. The present work extends our earlier results by the use of high field magnetometry. The experimental results are analyzed within the model based on spin waves theory. The theoretical results are able to reproduce the observed magnetic properties of the alloy for $x$$\,$$\leq$$\,$0.04 and prove, that carrier mediated Ruderman-Kittel-Kasuya-Yosida (RKKY)\cite{Ruderman54a, Kasuya56a, Yoshida57a} interaction was responsible for the observed magnetic states of the studied Ge$_{1\textrm{-}x}$Cr$_{x}$Te samples for $x$$\,$$<$$\,$0.04. The spinodal decomposition of Cr ions leads to saturation and decrease of the Curie temperature, $T_{C}$, for 0.04$\,$$<$$\,$$x$$\,$$<$$\,$0.061.

\section{Sample preparation and basic characterization}

\begin{figure}[t]
  \begin{center}
    \includegraphics[width = 0.5\textwidth, bb = 0 40 842 596]
    {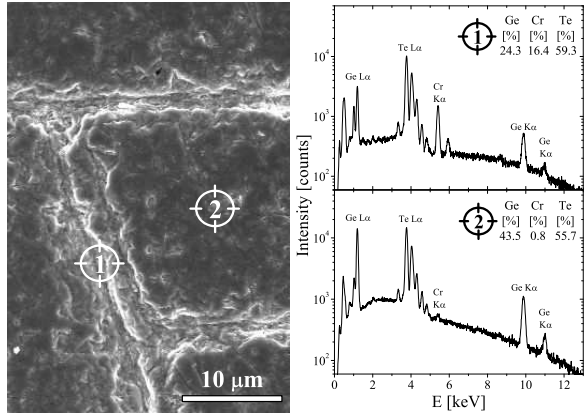}\\
  \end{center}
  \caption{\small SEM image of the surface of the selected Ge$_{0.941}$Cr$_{0.059}$Te crystal and the EDS spectra measured at selected spots of the sample.}
  \label{FigSemEds}
\end{figure}

Our Ge$_{1\textrm{-}x}$Cr$_{x}$Te crystals were grown using the modified Bridgman method. The growth procedure was modified similarly to the inclined crystallization front method, proposed by Aust and Chalmers for improving the quality of bulk alumina crystals.\cite{Aust58a} The presence of additional heating elements in the growth furnace created a radial temperature gradient in the growth furnace, changing the slope of the solid-liquid interface by about 15$^{\circ}$. The proposed modifications improved the crystal homogeneity and allowed the reduction of the number of crystal blocks in the as-grown ingot from a few down to a single one. \\ \indent The chemical composition of the samples was determined using the energy dispersive x-ray fluorescence technique (EDXRF).
\begin{table*}[t]
\caption{\label{tab:BasicCharact}
Results of basic characterization of Ge$_{1\textrm{-}x}$Cr$_{x}$Te samples performed at room temperature including chromium content $x$, lattice parameter $a$, angle of distortion $\alpha$, electrical resistivity $\rho_{xx}$, carrier concentration $n$ and mobility $\mu$.}
\begin{ruledtabular}
\begin{tabular}{cccccc}
 $x$$\,$$\pm$$\,$$\Delta$$x$ & $a$$\,$$\pm$$\,$$\Delta$$a$ (\AA) & $\alpha$$\,$$\pm$$\,$$\Delta$$\alpha$ (deg) & $\rho_{xx}$ (10$^{-3}$$\;$$\Omega$$\cdot$cm) & $n$ (10$^{20}$$\;$cm$^{-3}$) & $\mu$ (cm$^{2}$/(V$\cdot$s)) \\ \hline
 0.016$\,$$\pm$$\,$0.004 & 5.9847$\,$$\pm$$\,$0.0003 & 88.44$\,$$\pm$$\,$0.01  &  3.05$\,$$\pm$$\,$0.01  &  2.4$\,$$\pm$$\,$0.1  &  8.0$\,$$\pm$$\,$0.1  \\
 0.025$\,$$\pm$$\,$0.003 & 5.9845$\,$$\pm$$\,$0.0002 & 88.44$\,$$\pm$$\,$0.01  &  2.67$\,$$\pm$$\,$0.01  &  2.6$\,$$\pm$$\,$0.1  &  10.1$\,$$\pm$$\,$0.1  \\
 0.035$\,$$\pm$$\,$0.005 & 5.9837$\,$$\pm$$\,$0.0004 & 88.41$\,$$\pm$$\,$0.01  &  1.97$\,$$\pm$$\,$0.01  &  2.9$\,$$\pm$$\,$0.1  &  10.0$\,$$\pm$$\,$0.1  \\
 0.045$\,$$\pm$$\,$0.005 & 5.9829$\,$$\pm$$\,$0.0003 & 88.38$\,$$\pm$$\,$0.01  &  1.45$\,$$\pm$$\,$0.01  &  3.0$\,$$\pm$$\,$0.1  &  16.0$\,$$\pm$$\,$0.1  \\
 0.059$\,$$\pm$$\,$0.006 & 5.9829$\,$$\pm$$\,$0.0002 & 88.38$\,$$\pm$$\,$0.01  &  1.18$\,$$\pm$$\,$0.01  &  3.1$\,$$\pm$$\,$0.1  &  23.0$\,$$\pm$$\,$0.1  \\
 0.061$\,$$\pm$$\,$0.006 & 5.9820$\,$$\pm$$\,$0.0003 & 88.38$\,$$\pm$$\,$0.01  &  1.04$\,$$\pm$$\,$0.01  &  3.3$\,$$\pm$$\,$0.1  &  19.2$\,$$\pm$$\,$0.1  \\
\end{tabular}
\end{ruledtabular}
\end{table*}
The chemical composition of the ingots grown by this method changed continuously along the direction of crystal growth. To minimize the heterogeneity of the individual samples the as-grown crystals were cut perpendicular to the growth direction into thin slices each with a thickness of about 1$\;$mm. This ensured the heterogeneity of the individual slices to be relatively small with a maximum value of about $\Delta$$x$$\,$$\leq$$\,$0.006, where $\Delta x$ represents the variation of the amount of Cr in the slice. The results of the EDXRF measurements showed that the chemical composition of crystals changed continuously along the ingot from $x$$\,$$\simeq$$\,$0.012 up to 0.065. From all the crystal slices we selected the ones in which the relative chemical inhomogeneity was the smallest in the whole series (chemical composition of the samples is gathered in Table$\;$\ref{tab:BasicCharact}). \\ \indent The structural characterization of Ge$_{1\textrm{-}x}$Cr$_{x}$Te crystals was performed with the use of standard powder x-ray diffraction (XRD) technique. The XRD measurements were done at room temperature using Siemens D5000 diffractometer. The obtained diffraction patterns were then analyzed using Rietveld refinement method to determine the lattice parameters of each sample. The obtained results indicated that all the samples were single phase and crystallized in the NaCl structure with rhombohedral distortion in (111) direction. The lattice parameter $a$ as well as the angle of distortion $\alpha$ obtained for Ge$_{1\textrm{-}x}$Cr$_{x}$Te crystals are gathered in Table$\;$\ref{tab:BasicCharact}. The lattice parameters of the studied crystals are similar to the values for pure GeTe, i.e., $a$$\,$$=$$\,$5.98$\;$$\textrm{\AA}$ and $\alpha$$\,$$=$$\,$88.3$^{\circ}$.\cite{Galazka99a} The lattice parameter of Ge$_{1\textrm{-}x}$Cr$_{x}$Te samples is a decreasing function of chromium content $x$. The monotonic and close to a linear decrease of $a$ with $x$ is in accordance with the Vegard law, which is a clear signature that the Cr ions are incorporated to a large extent into cation positions in the GeTe crystal lattice. The $a$($x$) dependence can be fitted with a linear function $a$($x$)$\,$$=$$\,$5.9857$-$0.055$\cdot$$x$. The directional coefficient of the $a$($x$) function is about 3 times lower than the values reported in the literature for Ge$_{1\textrm{-}x}$Mn$_{x}$Te crystals.\cite{Fukuma03c} We believe that a large fraction of the chromium ions in the studied samples do not occupy the substitutional positions in the cation sublattice. We must therefore assume the presence of chromium ions in various charge states, which might have significant effects on the electrical and magnetic properties of the alloy. \\ \indent Scanning electron microscopy (SEM) was used to study chemical homogeneity of selected Ge$_{1\textrm{-}x}$Cr$_{x}$Te samples. Two techniques were used simultaneously: (i) field emission scanning electron microscopy with the use of Hitachi SU-70 Analytical UHR FE-SEM and (ii) Thermo Fisher NSS 312 energy dispersive x-ray spectrometer system (EDX) equipped with SDD-type detector. The SEM employed with this device allowed us to make microscopic pictures of the crystal surface. A series of measurements performed on different samples revealed the presence of a small concentration of pure Ge inclusions with  diameters typically around a few micrometers inside the samples. However, since there was no signature of chromium in the observed Ge inhomogeneities, we believe that they should not affect magnetic properties of the studied crystals. No signature of chromium clustering was observed (within the limits of accuracy of the method) in the case of Ge$_{1\textrm{-}x}$Cr$_{x}$Te samples with $x$$\,$$\leq$$\,$0.055. However, in the case of the samples with the highest studied compositions, i.e. $x$$\,$$>$$\,$0.055, some evidence of imperfect Cr dilution was observed (see Fig.$\;$\ref{FigSemEds}).
Detailed measurements showed a presence of Ge$_{1-x}$Cr$_{x}$Te accumulations with a chromium content varying in the range of $x$$\,$$\approx$$\,$0.40$\pm$0.05 and a diameter of around 20-30$\;$$\mu$m, diluted randomly inside the host lattice. It is highly probable, that these accumulations can affect the magnetic properties of Ge$_{1\textrm{-}x}$Cr$_{x}$Te alloys with $x$$\,$$>$$\,$0.055. \\ \indent The electrical properties of Ge$_{1\textrm{-}x}$Cr$_{x}$Te alloy were studied via resistivity and the Hall effect measurements performed at temperatures between 4.3 and 300$\;$K. We used a standard 6-contact dc current technique. The Hall effect measurements were done using a constant magnetic field $B$$\,$$=$$\,$1.4$\;$T. The results of electrical characterization of the samples obtained at room temperature are gathered in Table$\;$\ref{tab:BasicCharact}. The temperature dependent resistivity shows a behavior typical of a degenerate semiconductor, i.e. a metallic, nearly linear behavior of $\rho_{xx}$($T$) at $T$$\,$$>$$\,$30$\;$K and a plateau for $T$$\,$$<$$\,$30$\;$K. The data gathered in Table$\;$\ref{tab:BasicCharact} indicate a decrease in $\rho_{xx}$ value at room temperature with addition of chromium to the alloy. It might indicate that the carrier transport was strongly influenced by the addition of Cr ions into the alloy. However, more clear conclusions can be stated from the analysis of the Hall effect data (see Table$\;$\ref{tab:BasicCharact}). The results showed that all the studied Ge$_{1\textrm{-}x}$Cr$_{x}$Te crystals had a high $p$-type conductivity with relatively high carrier concentrations $n$$\,$$\approx$$\,$2...4$\times$10$^{20}$$\;$cm$^{-3}$. The data gathered in Table$\;$\ref{tab:BasicCharact} show that the Hall carrier concentration, $n$, is an increasing function of Cr content $x$. It is a clear signature, that some chromium ions do not substitute in cation lattice sites in GeTe host lattice, remaining in a different charge state than the Cr$^{2+}$ state. It is probable, that a small fraction of chromium ions remains as interstitial defects in the studied crystals. This may result in a change of a charge state of chromium and allows these ions to be electrically active. The Hall carrier mobility $\mu$, determined using a relation $\sigma_{xx}$$\,$$=$$\,$$e$$\cdot$$n$$\cdot$$\mu$, where $e$ is the elementary charge, is also an increasing function of chromium content $x$ up to $x$$\,$$=$$\,$0.059 (see Table$\;$\ref{tab:BasicCharact}). The simultaneous increase of the Hall carrier concentration $n$ and mobility $\mu$ with increasing chromium content $x$ must be of a complex origin. The main source of high $p$-type conductivity in IV-VI semiconductors is cation vacancies.\cite{Ure60a} We believe that the increasing $n(x)$ dependence was not originating from imperfect distribution of dopants in the semiconductor matrix. Probably, addition of chromium to the alloy changes slightly the thermodynamics of growth by inducing an increasing number of cation vacancies in the as grown crystals and thus leading to an increasing $n$($x$) function. On the other hand, the increasing $\mu$($x$) function must be connected with some weakening of the ionic scattering mechanism related to the negative charge state of Ge vacancies in GeTe. Thus, it is highly probable, that interstitial chromium ions are associated with Ge vacancies, screening their Coulomb potential and reducing their effective scattering cross-section. It seems that the addition of chromium ions to the GeTe matrix is an effective way to increase the mean free path of the conducting carriers, which can be a crucial factor when analyzing magnetic properties of the samples in question.

\section{Results and discussion}

The detailed studies of magnetic properties of Ge$_{1\textrm{-}x}$Cr$_{x}$Te crystals were performed including measurements of both static and dynamic magnetometry with the use of a LakeShore 7229 susceptometer/magnetometer and a Quantum Design superconducting quantum interference device (SQUID) MPMS XL-7 magnetometer system. The very same pieces of Ge$_{1\textrm{-}x}$Cr$_{x}$Te samples, that were previously characterized by means of magnetotransport measurements, with their electrical contacts removed, have been studied via several magnetometric methods. The SQUID magnetometer was used for determining the temperature dependencies of low-field magnetization and the Lake Shore 7229 dc magnetometer was used for measurements of high field isothermal magnetization curves.

\subsection{Low field results}

\begin{figure}[t]
  \begin{center}
    \includegraphics[width = 0.5\textwidth, bb = 0 40 595 520]
    {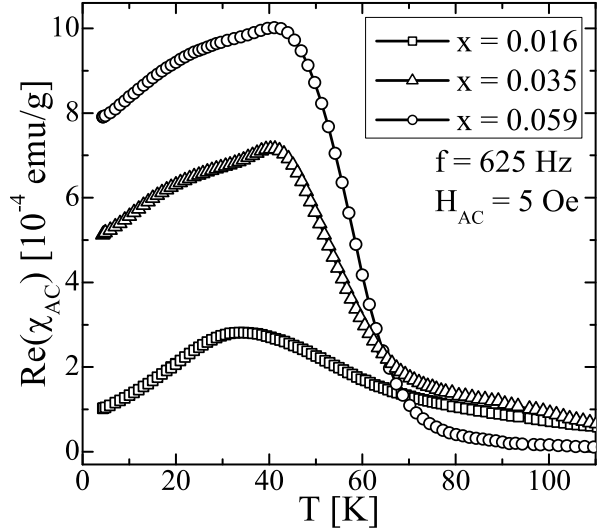}\\
  \end{center}
  \caption{\small The temperature dependence of the real part of the ac magnetic susceptibility $\chi_{AC}$ obtained for selected Ge$_{1\textrm{-}x}$Cr$_{x}$Te samples with different chemical composition $x$ (shown in legend).}
  \label{FigACSuscVsTemp}
\end{figure}

\begin{figure}[t]
  \begin{center}
    \includegraphics[width = 0.5\textwidth, bb = 0 30 590 520]
    {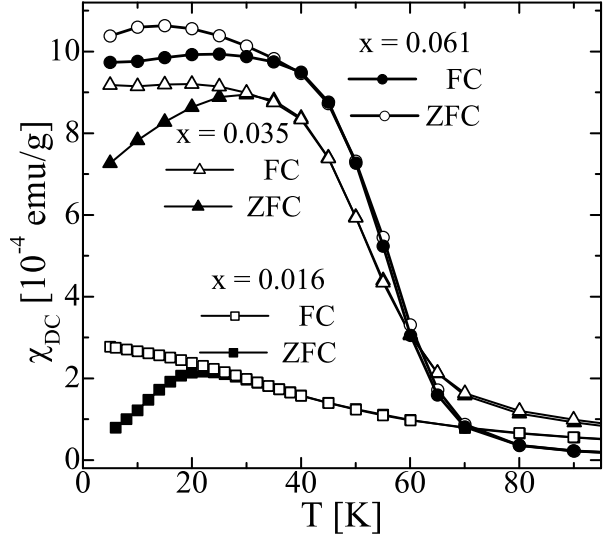}\\
  \end{center}
  \caption{\small The zero-field-cooled (ZFC) and field-cooled (FC) dc magnetic susceptibility as a function of temperature for selected Ge$_{1\textrm{-}x}$Cr$_{x}$Te crystals with different Cr content $x$ (see labels).}
  \label{FigMvTFCZFC}
\end{figure}

\begin{figure}[t]
  \begin{center}
    \includegraphics[width = 0.5\textwidth, bb = 0 30 590 520]
    {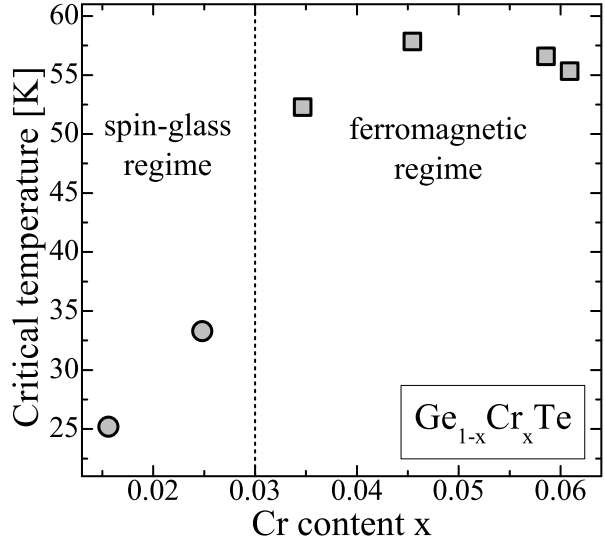}\\
  \end{center}
  \caption{\small The critical temperatures as a function of Cr content $x$ for the studied Ge$_{1\textrm{-}x}$Cr$_{x}$Te samples.}
  \label{FigTCTSGvx}
\end{figure}

The mutual inductance method employed in the LakeShore 7229 ac susceptometer system was used in order to measure the temperature dependence of both the real and imaginary parts of the ac susceptibility $\chi_{AC}$. The ac susceptibility was studied at temperatures in the range of 4.3$\,$$\leq$$\,$$T$$\,$$\leq$$\,$200$\;$K using an alternating magnetic field with amplitudes between 1$\,$$\leq$$\,$$H_{AC}$$\,$$\leq$$\,$20$\;$Oe and frequencies between 7$\,$$\leq$$\,$$f$$\,$$\leq$$\,$9980$\;$Hz. The real and imaginary parts of the susceptibility were measured as a function of temperature using a constant frequency $f$$\,$$=$$\,$625$\;$Hz and an applied magnetic field amplitude $H_{AC}$$\,$$=$$\,$5$\;$Oe. The results of the temperature dependent ac susceptibility measurements are presented in Fig.$\;$\ref{FigACSuscVsTemp}.
The Re($\chi_{AC}$(T)) curves for the samples with low chromium content, $x$$\,$$<$$\,$0.03, showed a broad peak with a maximum at temperatures lower than 35$\;$K. Peaks in the ac susceptibility shifted with frequency, as reported in Ref.$\;$\onlinecite{Kilanski11a}, and were convincingly identified as the appearance of a spin-glass freezing for $x$$\,$$<$$\,$0.03. However, a qualitatively different shape of the Re($\chi_{AC}$(T)) curves was observed for Ge$_{1\textrm{-}x}$Cr$_{x}$Te samples with $x$$\,$$>$$\,$0.03. A significant increase of the Re($\chi_{AC}$(T)) dependence as the temperature was lowered below 80$\;$K was followed by a maximum. We
believe that this is a signature that a well-defined magnetic phase transition occurred in these samples.  The increase was much smaller for samples with $x$$\,$$<$$\,$0.03. Moreover, in contrast to the Re($\chi_{AC}$(T)) dependence for samples with $x$$\,$$<$$\,$0.03, the magnetic susceptibility was not decreasing significantly with
temperatures lower than the temperature for maximum $\chi$. This is a signature that spontaneous magnetization appeared in the studied samples with $x$$\,$$>$$\,$0.03 at temperatures lower than 60$\;$K. In contrast to spin-glass samples, no signatures of frequency shifting of the maxima in the Re($\chi_{AC}$(T)) curves with increasing frequency of the ac magnetic field were observed. It is obvious, that a ferromagnetic order was observed in the case of Ge$_{1\textrm{-}x}$Cr$_{x}$Te samples with high chromium content $x$$\,$$>$$\,$0.03. However, more detailed studies of static magnetic properties need to be obtained in order to fully justify the above interpretation. \\ \indent The dc magnetometery was also used to study magnetization $M$ vs. $T$ of the Ge$_{1\textrm{-}x}$Cr$_{x}$Te crystals. At first, the temperature dependence of the magnetization was measured in the range of low magnetic field $H$$\,$$=$$\,$20$-$200$\;$Oe using a SQUID magnetometer. Measurements were performed over wide temperature range $T$$\,$$=$$\,$2$-$300$\;$K in two steps, i.e. with the sample cooled in the presence (FC) and the absence (ZFC) of the external magnetic field. The contribution of the sample holder was subtracted from the experimental data. The results showed that in the range of magnetic fields we used the isothermal magnetization $M$($H$) curves were nearly linear allowing us to calculate the dc magnetic susceptibility $\chi_{dc}$$\,$$=$$\,$$\partial M$/$\partial H$ and its changes with temperature for each Ge$_{1\textrm{-}x}$Cr$_{x}$Te sample. The temperature dependence of both ZFC and FC magnetic susceptibility for a few selected samples with different chemical content $x$ are gathered in Fig.$\,$\ref{FigMvTFCZFC}.
The ZFC $\chi_{dc}$($T$) curves showed different behavior in the case of Ge$_{1\textrm{-}x}$Cr$_{x}$Te samples with low ($x$$\,$$<$$\,$0.03) and high (0.03$\,$$<$$\,$$x$$\,$$<$$\,$0.062) chromium content, in agreement with the previously measured ac susceptibility. It should be noted, that in the case of spin-glass samples ($x$$\,$$<$$\,$0.03) the ZFC $\chi_{dc}$($T$) curve may be approximated to zero for $T$$\,$$\rightarrow$$\,$0. In the case of ferromagnetic samples ($x$$\,$$>$$\,$0.03) the separation between ZFC and FC curves was a decreasing function of chromium content. The features presented above are additional  proofs, consistent with the ac susceptibility results, that spin-glass ($x$$\,$$<$$\,$0.03) and ferromagnetic ($x$$\,$$>$$\,$0.03) magnetic order was observed in the case of the studied alloy. \\ \indent The critical behavior of the temperature dependencies of ZFC $M$($T$) curves is used to determine the Curie temperature, $T_{C}$, for each of the studied Ge$_{1\textrm{-}x}$Cr$_{x}$Te samples. We determined the values of $T_{C}$ from the position of the inflection point on the low-field magnetization curves, $M$($T$). The values of the estimated critical temperatures for all the studied Ge$_{1\textrm{-}x}$Cr$_{x}$Te samples, including our earlier estimations of the spin-glass transition temperatures $T_{SG}$ for samples with $x$$\,$$<$$\,$0.026 (see Ref.$\;$\onlinecite{Kilanski11a}) are gathered in Fig.$\;$\ref{FigTCTSGvx}. It should be noted, that the values of both $T_{SG}$ and $T_{C}$ estimated from the dynamic susceptibility results were close to those estimated with the use of static magnetization results.
The results presented in Fig.$\;$\ref{FigTCTSGvx} show clearly the monotonic increase of the critical temperature of the alloy with the increase in the chromium concentration for 0.016$\,$$\leq$$\,$$x$$\,$$\leq$$\,$0.045. The monotonic increase of the critical temperature with $x$ for the samples showing both spin-glasslike and ferromagnetic ordering substantiates the assumption that both types of magnetic order originated from a single type of long range RKKY magnetic interactions. Long range RKKY magnetic interactions promote an appearance of a spin-glass state at low paramagnetic ions content. It is connected with the similar probability of finding two spins with both positive and negative sign of the exchange constant at low dilution limit. Due to this probability, the appearance of a spin-glass state is expected for 0.016$\,$$\leq$$\,$$x$$\,$$\leq$$\,$0.025. The increase of chromium amount increases the probability of magnetic interactions with positive sign of the exchange constant. This results in an appearance of a ferromagnetic state in Ge$_{1\textrm{-}x}$Cr$_{x}$Te samples for $x$$\,$$\geq$$\,$0.035.

\subsection{High field magnetization}

\begin{figure}[t]
  \begin{center}
    \includegraphics[width = 0.5\textwidth, bb = 0 40 590 520]
    {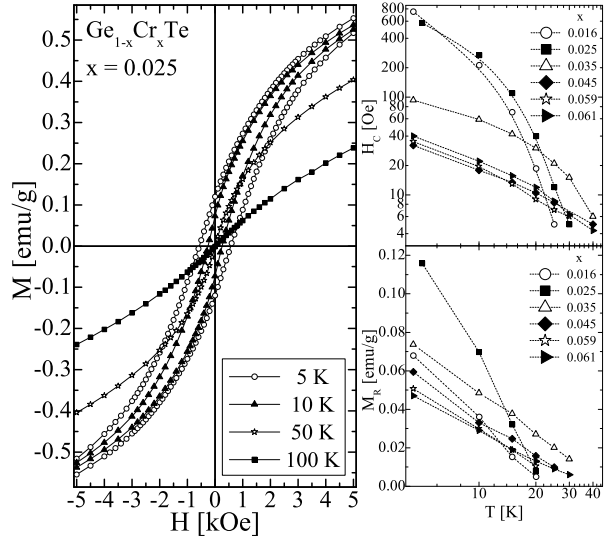}\\
  \end{center}
  \caption{\small Magnetic hysteresis curves obtained at different temperatures for the Ge$_{0.975}$Cr$_{0.025}$Te sample (left panel) together with temperature dependence of coercive field $H_{C}$ and spontaneous magnetization $M_{R}$ (right panel).}
  \label{FigMHist}
\end{figure}

\begin{figure}[t]
  \begin{center}
    \includegraphics[width = 0.5\textwidth, bb = 0 40 595 520]
    {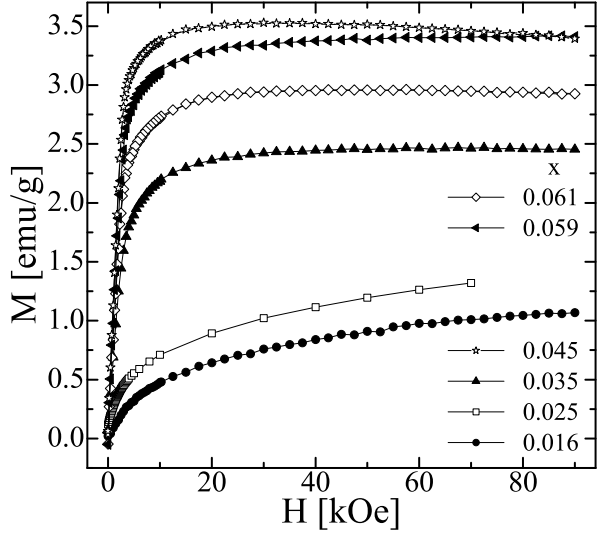}\\
  \end{center}
  \caption{\small High field magnetization curves obtained at $T$$\,$$=$$\,$4.5$\;$K for the studied Ge$_{0.975}$Cr$_{0.025}$Te samples with different chromium content $x$.}
  \label{FigMHHighField}
\end{figure}

The magnetic behavior of the Ge$_{1\textrm{-}x}$Cr$_{x}$Te samples was investigated with the use of magnetization $M$ measurements in the presence of a dc magnetic field up to $H$$\,$$=$$\,$90$\;$kOe. We used the Weiss extraction method included in the LakeShore 7229 magnetometer system. At first, detailed isothermal magnetic hysteresis loops were measured at selected temperatures $T$$\,$$<$$\,$200$\;$K. A clear magnetic hysteresis was observed in all the studied crystals at temperatures lower than either $T_{SG}$ or $T_{C}$. It should be emphasized, that a hysteretic behavior was observed in both the spin-glass and the ferromagnetic regime of the studied alloy (see Fig.$\;$\ref{FigMHist}).
The appearance of a well defined hysteretic behavior in a spin-glass system is a signature that a strong magnetic disorder is causing significant magnetic frustration in the samples with $x$$\,$$<$$\,$0.03. Such a feature was observed in many representatives of metallic spin-glasses such as AuFe with 8 at.\% Fe (Ref.$\;$\onlinecite{Prejean80a}) or NiMn with 21 at.\% Mn (Ref.$\;$\onlinecite{Senoussi84a}). \\ \indent A series of magnetic hysteresis loops obtained at temperatures $T$$\,$$<$$\,$100$\;$K was used to determine the temperature dependencies of both coercive field $H_{C}$ and spontaneous magnetization $M_{R}$ (see the right panel of Fig.$\;$\ref{FigMHist}). The $H_{C}$ observed at $T$$\,$$=$$\,$4.5$\;$K changed monotonically as a function of $x$ indicating that large changes in the domain structure of the studied system occurred while changing its chemical composition. It is especially visible in the case of the $H_{C}$($x$) dependence, where we observed differences higher than an order of magnitude; $H_{C}$ depended on the temperature and at 4.5$\;$K was 800$\;$Oe for $x$$\,$$=$$\,$0.016 and 40$\;$Oe for $x$$\,$$=$$\,$0.061. The trend in the $M_{R}$ with $x$ had more complicated character first increasing for $x$$\,$$\leq$$\,$0.025 and then decreasing as a function of $x$. \\ \indent The high-field isothermal magnetization curves were studied in the case of all the Ge$_{1\textrm{-}x}$Cr$_{x}$Te samples at $H$$\,$$\leq$$\,$90$\;$kOe and $T$$\,$$<$$\,$200$\;$K. In order to compare the results we plotted $M$($H$) curves measured at the lowest measurement temperature $T$$\,$$=$$\,$4.5$\;$K for the samples having different chemical composition (see Fig.$\;$\ref{FigMHHighField}).
As we can see, the magnetization curves have different shapes for samples showing spin-glass ordering ($x$$\,$$<$$\,$0.03) than for the ferromagnetic ones ($x$$\,$$>$$\,$0.03). The $M$($H$) curves in the case of the spin-glass samples saturated slowly and even at high magnetic fields $H$$\,$$=$$\,$90$\;$kOe, did not reach the saturation. In case of ferromagnetic crystals the magnetization reached a saturation value for magnetic fields below 20$\;$kOe. It should be noted, that for the samples with $x$$\,$$>$$\,$0.03 the magnetization is slowly decreasing as a function of magnetic field for $H$$\,$$>$$\,$30$\;$kOe. This is a diamagnetic contribution from the GeTe lattice. \\ \indent The saturation magnetization $M_{S}$($T$$\,$$=$$\,$4.5$\;$K) was an increasing function of Cr content for $x$$\,$$<$$\,$0.05, and showed a small decrease for higher $x$. The amount of the magnetically active Cr ions in the material can be determined from the value of the saturation magnetization, $M_{S}$ observed at $T$$\,$$=$$\,$4.5$\;$K. We use the typical formula describing the saturation magnetization for the calculation of the amount of Cr in the studied compound, $\bar{x}$$\,$$=$$\,$($M_{S}$$m_{u}$)/($N_{Av}$$g$$J$$\mu_{B}$), where $m_{u}$ is the particle mass of the Ge$_{1\textrm{-}x}$Cr$_{x}$Te alloy, $N_{Av}$ is the Avogadro constant, $g$ is the effective spin splitting factor, $J$$\,$$=$$\,$$S$$\,$$=$$\,$2 is the total momentum of the Cr$^{2+}$ ion, and $\mu_{B}$ is the Bohr magneton. The $\bar{x}$ estimation assumes that the spin-component of the momentum $S$$\,$$=$$\,$2 in the GeTe crystal field. The obtained values of $\bar{x}$ are about 2-times lower than the values obtained using the EDXRF method indicating, that half of the Cr ions in the alloy are magnetically inactive. It is also possible that some Cr ions are in the Cr$^{3+}$ state, with $J$$\,$$=$$\,$$S$$\,$$=$$\,$3/2. Let us also mention that for purely random, uncorrelated distribution of substitutional Cr ions, some fraction of them forms clusters, in which nearest-neighbor ions couple antiferromagnetically due to superexchange mechanism.

\subsection{Estimation of the exchange integral $J_{pd}$}

\begin{figure}[t]
  \begin{center}
    \includegraphics[width = 0.5\textwidth, bb = 0 40 600 540]
    {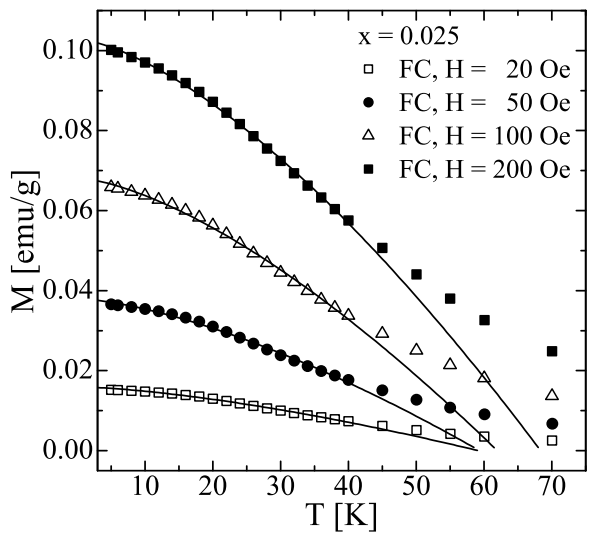}\\
  \end{center}
  \caption{\small The temperature dependence of the FC magnetization obtained experimentally (symbols) and calculated (lines) for different magnetic field  values (see legend) for the Ge$_{0.975}$Cr$_{0.025}$Te sample.}
  \label{FigMTTheor}
\end{figure}

In order to gain more insight into the physical mechanism behind the magnetic ordering in the samples with low Cr content $x$$\,$$=$$\,$0.016 and 0.025 (where the Cr distribution is homogeneous and no spinodal decomposition was observed), SQUID measurements of the temperature dependence of magnetization in weak external fields of 20, 50, 100 and 200 Oe were performed. The results for the sample with $x$$\,$$=$$\,$0.025 are presented in Fig.~\ref{FigMTTheor}. Our interest was particularly focused on examining whether the Bloch law for magnetization is fulfilled in the form $M\left(T\right) = M\left(0\right)\left[1-\left(T/T_{SW}\right)^{3/2}\right]$. The parameter $T_{SW}$ is related to the spin-wave stiffness $D$ by $T_{SW}=\frac{4\pi D}{k_{B} a^{2}} \left(\frac{4J}{\zeta\left(3/2\right)}\right)^{2/3}$, where $a$ is the lattice constant and $\zeta\left(3/2\right)\simeq 2.612$ is the appropriate Riemann zeta function, while $J$=2 is the ionic angular momentum.\cite{Kilanski10a, Cochrane74a} Spin wave stiffness is defined by the relation between spin wave energy $E$ and wavevector $q$ of the form $E=Dq^2$. The results of fitting the formula to the low-temperature part of the experimental data (excluding the range showing a 'magnetization tail') are shown in Fig.~\ref{FigMTTheor} with solid lines.
It can be observed that the temperature behavior of magnetization follows very well the $T^{3/2}$-dependence, for all the values of low external field. What is important is that the characteristic temperatures $T_{SW}$ determined from the fittings exhibit rather weak dependence on the external field, especially for $H$$\,$$=$$\,$20 and 50$\;$Oe. Therefore, we accept the average value of $T_{SW}$ determined for these two weakest external fields as a measure of the spin-wave stiffness. For $x$$\,$$=$$\,$0.016 we obtained $T_{SW}$$\,$$=$$\,$53.5$\;$K (which corresponds to $D$$\,$$=$$\,$6.2$\;$meV$\cdot$$\textrm{\AA}^{2}$), while for $x$$\,$$=$$\,$0.025 $T_{SW}$$\,$$=$$\,$59.3$\;$K (yielding $D$$\,$$=$$\,$6.9$\;$meV$\cdot$$\textrm{\AA}^{2}$). \\ \indent We attempted to apply the RKKY model to reproduce the experimental values of $T_{SW}$. Within the harmonic spin-wave and virtual crystal approximation, $D=\frac{1}{6}\overline{x}J\sum_{i}^{}{J_{RKKY}\left(R_{0i}\right)R^{2}_{0i}}$, where summation is performed over all the lattice sites $i$, i.e. over all possible positions of substitutional magnetic impurity ions, relative to a selected lattice site denoted by 0.\cite{Kilanski10a} Let us note that instead of the EDXRF-measured Cr content $x$ we use the $\overline{x}$ values, estimated from the low-temperature high-field magnetization (see previous section). We expect that only this magnetically active fraction of Cr ions participates in spin wave excitations in the system. The values amount to $\overline{x}=0.0096$ for the sample with $x=0.016$ and $\overline{x}=0.013$ for $x=0.025$, respectively. The indirect RKKY exchange integral between a pair of magnetic ions at lattice sites 0 and $i$ with the relative distance $R_{0i}$ equals \cite{Ruderman54a, Kasuya56a, Yoshida57a}
\begin{widetext}
\begin{equation}\label{Eq03}
J_{RKKY}(R_{0i})=N_{V}\frac{m^{*}J_{pd}^{2}a^{6}k_{F}^{4}}{32\pi^{3}\hbar^{2}}
\times \frac{\sin(2k_{F}R_{0i})-2k_{F}R_{0i}\cos(2k_{F}R_{0i})}{(2k_{F}R_{0i})^{4}}\,\exp(-\frac{R_{0i}}{\lambda}),
\end{equation}
\end{widetext}
where $m^{*}$ is the effective mass of the carriers residing in $N_{v}=4$ valleys in the valence band, $J_{pd}$ is the exchange integral between charge carriers and Cr ions, while $k_{F} = (3 \pi^{2} n / N_{V})$ denotes the Fermi wave vector. \\ \indent In order to account for the presence of a disorder and a finite mean free path of the charge carriers mediating the RKKY coupling we assume an exponential damping with a length scale $\lambda$. We estimated the values of $\lambda$ on the basis of the Drude model of conduction taking into account the experimentally observed low-temperature values of the Hall carrier concentration and mobility. Knowing the values of charge carriers concentration for $T\to 0$ as well as the mobility $\mu$, we can calculate the mean free path as $\lambda=\left(\hbar k_{\rm F}\mu\right)/e$. For the sample with $x=0.016$ we obtained $\lambda=$9.5 \AA, while for $x=0.025$ we obtained $\lambda=12.5$ \AA. We use these estimates for the RKKY-based model of spin-wave stiffness. \\ \indent Under such assumptions, we are able to reproduce the experimental values of $T_{SW}$ using $J_{pd}=$0.88$\pm$0.05 eV for $x$$\,$$=$$\,$0.016 and $J_{pd}$$\,$$=$$\,$0.77$\pm$0.05 eV  for $x$$\,$$=$$\,$0.025. The obtained estimates are close to each other, within the experimental uncertainty, including the uncertainty in determination of the magnetically active Cr concentration in the samples, $\bar{x}$.

\section{Summary and conclusions}

We presented detailed studies of magnetic properties of Ge$_{1\textrm{-}x}$Cr$_{x}$Te bulk crystals grown using the modified Bridgman method with chromium content changing in the range of 0.016$\,$$\leq$$\,$$x$$\,$$\leq$$\,$0.061. The X-ray diffraction studies revealed that a Vegard law was fulfilled in the alloy indicating a proper solubility of chromium in the crystal lattice. All the studied samples were $p$-type semiconductors with high carrier concentrations $n$$\,$$\approx$$\,$2.4$...$3.3$\times$10$^{20}$$\;$cm$^{-3}$ and mobilities $\mu$$\,$$\approx$$\,$8$...$23$\;$cm$^{2}$/(Vs). \\ \indent The magnetic properties of the Ge$_{1\textrm{-}x}$Cr$_{x}$Te are composition dependent. The presence of spin-glass and ferromagnetic phase was observed at $T$$\,$$<$$\,$60$\;$K, for the samples with $x$$\,$$<$$\,$0.03 and $x$$\,$$\geq$$\,$0.03, respectively. The RKKY interactions are found to be responsible for the observed magnetic ordering for $x$$\,$$<$$\,$0.045, while for $x$$\,$$\geq$$\,$0.045 the spinodal decomposition of Cr ions leads to saturation and decrease of the Curie temperature, $T_{C}$, with increasing $x$. The hysteretic behavior was observed both for spin-glass and ferromagnetic samples at $T$$\,$$<$$\,$50$\;$K with strong composition dependencies of the coercive field $H_{C}$. The amount of magnetically active Cr ions, $\bar{x}$, estimated from the value of the saturation magnetization $M_{S}$, was about 2-times lower than the value of the Cr content $x$ obtained using the EDXRF method. The calculations based on spin waves model reproduced the observed magnetic properties of the alloy for $x$$\,$$<$$\,$0.04 and proved, that carrier mediated RKKY interaction is responsible for the observed magnetic states. The value of the Cr-hole exchange integral $J_{pd}$, estimated via fitting of the experimental results with the theoretical model was about 0.77$\;$eV for $x$$\,$$=$$\,$0.025 and 0.88$\;$eV for $x$$\,$$=$$\,$0.016.

\section{Acknowledgments}

\noindent Scientific work was financed from funds for science in 2009-2013, under the project no. IP2010017770 granted by Ministry of Science and Higher Education of Poland and project no. N$\,$N202$\,$166840 granted by National Center for Science of Poland. This work has been partly supported by the Foundation for Polish Science under the project POMOST/2011-4/2. \\ \indent This work has been partly supported by Polish Ministry of Science and Higher Education by a special purpose grant to fund the research and development activities and tasks associated with them, serving the development of young scientists and graduate students.

\end{document}